\documentclass[a4paper,12pt,epsf]{article}
\usepackage{amsmath}
\usepackage[psamsfonts]{amssymb}
\usepackage{rsfs}
\usepackage{bm}
\usepackage{eucal}

\usepackage{cite}
\usepackage{graphicx}
\usepackage{tabularx}
\usepackage{layout}
\usepackage{color}

\begin{document} 

\begin{titlepage}

\hrule 
\leftline{}
\leftline{Chiba Univ. Preprint
          \hfill   \hbox{\bf CHIBA-EP-129}}
\leftline{\hfill   \hbox{hep-th/0105299}}
\leftline{\hfill   \hbox{May 2001}}
\vskip 5pt
\hrule 
\vskip 1.0cm
\centerline{\large\bf 
Vacuum condensate of mass dimension 2 
} 
\vskip 0.5cm
\centerline{\large\bf  
as the origin of mass gap and quark confinement  
}
\vskip 0.5cm
\centerline{\large\bf  
}

\vskip 0.5cm

\centerline{{\bf 
Kei-Ichi Kondo$^{\dagger}$,
}}  
\vskip 0.5cm
\begin{description}
\item[]{\it \centerline{ 
Department of Physics, Faculty of Science, 
Chiba University,  Chiba 263-8522, Japan}
}
\end{description}

\centerline{{\bf Abstract}}
 
We propose a vacuum condensate of mass dimension 2 consisting of gluons and ghosts in the framework of the manifestly covariant gauge fixing of the $SU(N)$ Yang-Mills theory. This quantity is both BRST and anti-BRST invariant for any gauge.  It includes the ghost condensation $C^a \bar{C}^a$ proposed first in the modified Maximal Abelian gauge and reduces to the gluon condensates $(\mathscr{A}_\mu)^2$ of mass dimension 2 proposed recently in the Landau gauge of the Lorentz gauge fixing.  
The vacuum condensate of dimensions 2 can provide the effective mass for gluons and ghosts.  
The possible existence of such condensations is demonstrated by calculating the operator product expansion of the gluon and ghost propagators in both gauges.
Its implications to quark confinement are also discussed in consistent with the previous works.


\vskip 0.2cm
${}^\dagger$ 
  E-mail:  {\tt kondo@cuphd.nd.chiba-u.ac.jp}

\vskip 0.2cm  
\hrule  

\par 
\par\noindent




\pagenumbering{roman}

\vskip 0.5cm  

\end{titlepage}


\pagenumbering{arabic}

\section{\label{sec:intro}Introduction}

It is widely believed that the QCD vacuum is characterized by the vacuum condensate and that the vacuum condensate is a key object to understand the non-perturbative dynamics of QCD.  The most famous example is the quark condensate for light quarks ($u$ or $d$):
\begin{equation}
  \langle 0| \bar{q}q |0 \rangle \not= 0 .
\end{equation}
The quark condensate has the canonical dimension 3.
A nonvanishing value of the quark condensate signals the spontaneous breaking of chiral symmetry.  
\par
Now we raise a question as to whether there is a vacuum condensate which is able to signal the quark confinement.  
Indeed, in the gluon sector, the gluon condensate of dimension 4:
\begin{equation}
 \langle 0| \alpha_s (\mathscr{F}_{\mu\nu}^A)^2 |0 \rangle \not= 0 ,
\label{gluoncond4}
\end{equation}
has played the crucial role for suggesting the existence of nonperturbative vacuum in the framework of the QCD sum rule\cite{SVZ79}.  Here the Capital index $A$ runs over $A=1, \cdots, N^2-1$ for the gauge group $G=SU(N)$.  
The gluon condensate of dimension 4 (\ref{gluoncond4}) has something to do with the quark confinement, since quark confinement is believed to be caused by a nonperturbative dynamics among gluons.
\par
Recently,  a novel type of a vacuum condensate of dimension 2, i.e., ghost  condensation 
(more precisely, ghost--anti-ghost condensation)%
\footnote{
If we allow the spontaneous breaking of the ghost number $Q_C$ to take place, the ghost--ghost and anti-ghost--anti-ghost condensations are not prohibited\cite{KS00}.  In this paper, we assume 
$Q_C |0 \rangle=0$.
}
 \cite{Schaden99,KS00} of the type,
\begin{equation}
  \langle 0| ig^2 \bar{C}^a C^a |0 \rangle \not= 0 ,
\label{ghostcond}
\end{equation}
has been proposed as a signal of quark confinement \cite{Kondo00,KondoI}.  Here the italic index $a$ runs over the same range as the Capital index $A$ but the indices $i$ of the Cartan subalgebra (We call the field with the index $a$ the off-diagonal component, while the field with the index $i$ of the Cartan subalgebra the diagonal component hereafter.  For $G=SU(2)$, we choose $a=1,2$ and $i=3$.)
The unequal treatment of the off-diagonal and diagonal components is a consequence of the Maximal Abelian gauge (MA) gauge.
The MA gauge is a covariant (partial) gauge fixing which restrict the original non-Abelian gauge group $G=SU(N)$ to the maximal torus group $H=U(1)^{N-1}$.
After the MA gauge fixing for the off-diagonal gluon components $A_\mu^a$, the theory has the residual $H$ local gauge symmetry.  
In particular, we have adopted a modified version of the MA gauge (the modified MA gauge \cite{KondoII}) which has several advantages compared with the naive MA gauge \cite{KondoI}.  
The MA gauge is a unique gauge which is able to realize the dual superconductivity \cite{dualsuper,tHooft81} of the QCD vacuum in the sense that it leads to the infrared Abelian dominance \cite{EI82,SY90} and that it allows us to derive the dual Ginzburg-Landau theory and the other equivalent theories, e.g., antisymmetric tensor theory, the magnetic monopole theory, a confining string theory as low-energy effective theories of QCD, see  \cite{Kondo00}.
However, the analyses carried out so far for the ghost condensation were in the rather heuristic level.  

\par
Soon after the works \cite{Schaden99,KS00} (but independently), the existence of another vacuum condensate of dimension 2:
\begin{equation}
 \langle 0| \mathscr{A}_\mu^A \mathscr{A}_\mu^A |0 \rangle \not= 0 ,
\label{gluoncond2}
\end{equation}
has been stressed by Boucaud et al.\cite{Boucaud00} and Gubarev, Stodolsky and Zakhaharov\cite{GSZ01}.
Their studies are based on the conventional covariant gauge fixing of the Lorentz type, especially they have chosen the Landau gauge $\alpha=0$ for the gauge fixing parameter.
The physical meaning of the gauge non-invariant condensate
$\langle (\mathscr{A}_\mu^A)^2 \rangle$ has been nicely explained by Gubarev and Zakharov \cite{GZ01} and a close connection to the topological defects, i.e., magnetic monopole has been suggested. 
Although the quantity $\langle (\mathscr{A}_\mu^A)^2 \rangle$ is gauge non-invariant,  
the minimal value of the integrated squared potential $\mathscr{A}^2_{min}$ has a definite physical meaning. This was demonstrated by the numerical simulation of {\it compact} U(1) lattice gauge theory where $\mathscr{A}^2_{min}$ signals the quark confinement in the sense that  it is non-zero only in the strong coupling region $e>e_c$ where quark confinement occurs due to monopole condensation.
Monopole condensation is believed to be an essential mechanism of quark confinement also in QCD.  
In fact, the magnetic monopole dominance in the MA gauge for quark confinement (i.e., saturation of the string tension by the magnetic monopole configuration) has been confirmed by numerical simulation on a lattice \cite{monopole}.
Therefore, these results strongly suggest the importance of dimension 2 vacuum condensates in understanding the quark confinement also in QCD.  

\par
One of the purposes of this paper is to announce the results of more systematic investigations (along the idea \cite{KS00}) performed based on the operator product expansion (OPE)\cite{Wilson69} and the effective action derived from the Wilson renormalization group (RG)\cite{WK74},
although the details will be published in a subsequent long paper\cite{KIMS01}.  
  The results of these analyses show that in the MA gauge
another vacuum condensate of dimension 2, i.e.,
\begin{equation}
  \langle 0| g^2 A_\mu^a A_\mu^a |0 \rangle \not= 0 ,
\label{gluoncond}
\end{equation}
together with (\ref{ghostcond}) 
must exist in the Yang-Mills theory and that both condensates can be responsible for quark confinement in QCD.  
In fact, we can show that the existence of two kinds of the vacuum condensates of dimensions 2, i.e., (\ref{ghostcond}) and (\ref{gluoncond}), are consistent with the OPE and RG. In fact, the RG analysis show that the linear combinations of two condensates behaves as the nonvanishing masses for the off-diagonal gluons and off-diagonal ghost (and anti-ghost) and that the masses are the invariants under the RG.  This result strongly supports the infrared Abelian dominance which has been confirmed by numerical simulations on a lattice in the last decade.
Thus both condensates constitute basic ingredients for deriving the dual superconductor picture of the QCD vacuum, as demonstrated in a previous paper\cite{Kondo00}.
\par
\par
As is well known, the Yang-Mills theory with the gauge fixing has  a global symmetry, i.e., the Becchi-Rouet-Stora-Tyutin (BRST) symmetry and the anti-BRST symmetry \cite{BRST,antiBRST} which play the role of the substitute of the gauge symmetry which was lost by the gauge fixing procedure.
The original Yang-Mills Lagrangian does not involve both the gluon bilinear term, 
$\mathscr{A}_\mu \cdot \mathscr{A}_\mu$ 
and the ghost bilinear term
$\bar{\mathscr{C}} \cdot \mathscr{C}$, since the
inclusion of such terms in the Lagrangian breaks the gauge invariance and the BRST invariance.
Therefore, they should be generated dynamically as a quantum effect.
The vacuum condensates of dimension 2  (which is gauge dependent) does not appear in the OPE of the gauge invariant quantity, e.g., the spectral function 
$\Pi(Q^2):=i \int d^4x e^{iqx} \langle 0| T[j(x),j(0)]|0 \rangle ,
$ where $q^2=-Q^2$ and $j(x)$ are local currents constructed on the quark and gluon fields.  
This is a reason why the dimension 2 vacuum condensate has not been seriously discussed so far.
However, the OPE of the gauge non-invariant quantity, e.g., the propagator and the vertex function, can involve the gauge (parameter) dependent and gauge non-invariant vacuum condensate of dimension 2.
Each composite operator providing the vacuum condensate, (\ref{ghostcond}) and (\ref{gluoncond}), is not BRST invariant. 
In view of these, we guess that two types of vacuum condensates of dimension 2 might be related to each other so that an appropriate combination of two condensates may give a BRST invariant quantity.  
Another purpose of this paper is to show that this is indeed the case:  
we propose a BRST and anti-BRST invariant combination of dimension 2 vacuum condensates,%
\footnote{
Note that there is a freedom of choosing a overall factor.  Therefore, only the ratio of two coefficients in the linear combination (\ref{combi}) is fixed from the requirement of the BRST and anti-BRST invariance. 
}
\begin{equation}
 \mathcal{O} := {1 \over \Omega^{(D)}} \int d^D x \ \text{tr} \left[ 
 {1 \over 2} \mathscr{A}_\mu(x) \cdot \mathscr{A}_\mu(x) - \alpha i \mathscr{C}(x) \cdot \bar{\mathscr{C}}(x) 
\right] ,
\label{combi}
\end{equation}
where $\Omega^{(D)}$ is the volume of the $D$-dimensional spacetime.
Here the trace is taken over the broken generators of the Lie algebra of the original group $G$, i.e, $G$ itself for the Lorentz gauge and $G/H$ for the partial (MA) gauge (i.e., off-diagonal components).  
The BRST invariance of $\mathcal{O}$ is shown in Appendix.
The existence of a linear combination of a number of operators with the same mass dimension and the same symmetry is very natural from the viewpoint of the operator mixing.  
Moreover, we suggest that it is an order parameter for quark confinement in the general formulation of the Yang-Mills theory in the manifestly covariant gauge fixing.  
This criterion should be compared with the well-known gauge-invariant criterion of quark confinement, i.e., the area law decay of the Wilson loop.
It is clear that $\mathcal{O}$ reduces to the $\mathscr{A}^2_{min}$ in the Landau gauge $\alpha=0$.  Therefore, the Landau gauge turns out to be a rather special case in which we do not need to consider the ghost condensation.
\par
The expectation value of $\mathcal{O}$ in the translational invariant vacuum yields
$
 \langle \mathcal{O} \rangle 
= \langle \text{tr} \left[ {1 \over 2} \mathscr{A}_\mu(0) \cdot \mathscr{A}_\mu(0) - \alpha i \mathscr{C}(0) \cdot \bar{\mathscr{C}}(0)\right]  \rangle .
$
In the effective potential approach, this value is obtained as the stationary point $(\xi_0, \eta_0)$ which gives the minimum of the effective potential $U(\xi,\eta)$
written in terms of the auxiliary fields $\xi$ and $\eta$ corresponding to the composite operators, 
$\text{tr} \left[ {1 \over 2} \mathscr{A}_\mu \cdot \mathscr{A}_\mu \right]$ 
and 
$\text{tr} \left[ i \mathscr{C} \cdot \bar{\mathscr{C}} \right]$, respectively. 
In the MA gauge, it is shown that an additional ghost condensation 
$\langle i \epsilon^{ab}C^a \bar{C}^b \rangle$ is necessary and hence we need to consider three composite operators:
$A_\mu^a A_\mu^a$, $iC^a \bar{C}^a$ and $i \epsilon^{ab}C^a \bar{C}^b$.
Hence the analysis becomes rather complicated, for details see \cite{KIMS01}.

\section{OPE of the gluon and ghost propagators}

\subsection{The Lorentz gauge}
First, we consider the question:  What is the most general Lagrangian that is a local function of the fields, $\mathscr{A}_\mu^A, \mathscr{B}^A, \mathscr{C}^A, \bar{\mathscr{C}}^A$ and satisfies the following conditions?
(1) of mass dimension 4, 
(2) Lorentz invariant,
(3a) BRST invariant,
(3b) anti-BRST invariant,
(4) Hermitian,
(5) zero ghost number 
(6) global gauge invariant, 
(7) (multiplicative) renormalizable.
Here it is implicitly assumed that the Lagrangian is restricted to a polynomials of the fields, and that there is no higher derivative terms, since there is no intrinsic mass scale in the Yang-Mills theory.
The answer was given by Baulieu and Thierry-Mieg \cite{BT82,Baulieu85}:  The Lagrangian is given by
\begin{equation}
\begin{align}
  \mathscr{L}^{tot} &= -{1 \over 4} \alpha_1 \mathscr{F}_{\mu\nu}^A \mathscr{F}^{\mu\nu}{}^A 
+ \alpha_2 \epsilon_{\mu\nu\rho\sigma}  \mathscr{F}^{\mu\nu}{}^A \mathscr{F}^{\rho\sigma}{}^A 
\nonumber\\
& + i \bm{\delta}_B \bar{\bm{\delta}}_B \left( \alpha_3 \mathscr{A}_\mu^A \mathscr{A}^\mu{}^A + \alpha_4 \mathscr{C}^A \bar{\mathscr{C}}^A \right) 
+ {\alpha' \over 2} \mathscr{B}^A \mathscr{B}^A ,
\end{align}
\end{equation}
where $\alpha_i(i=1,2,3,4)$ are arbitrary constant, and $\bm{\delta}_B$ and $\bar{\bm{\delta}}_B$ are the BRST and anti-BRST transformations.
The first term is the Yang-Mills Lagrangian, the second term is the topological term which is not discussed in this paper.  The first and the second terms are gauge invariant.  On the other hand, the third and the fourth terms are identified with the gauge fixing term, since they break the gauge invariance of the Lagrangian.  After the rescaling of the parameters into the field redefinitions, the above result leads to the GF+FP term \cite{DJ82}:
\begin{equation}
  \mathscr{L}_{GF+FP} = i \bm{\delta}_B \bar{\bm{\delta}}_B \left( {1 \over 2} \mathscr{A}_\mu^A \mathscr{A}^\mu{}^A -{\alpha \over 2}i \mathscr{C}^A \bar{\mathscr{C}}^A \right) 
+ {\alpha' \over 2} \mathscr{B}^A \mathscr{B}^A .
\label{GFglobal}
\end{equation}
The final term is allowed for the renormalizability of the total Lagrangian and is written in the BRST exact or anti-BRST exact form,
$\mathscr{B}^A \mathscr{B}^A=-i\bm{\delta}_B(\bar{\mathscr{C}}^A \mathscr{B}^A)=i\bar{\bm{\delta}}_B(\mathscr{C}^A\mathscr{B}^A)$.  By performing the BRST and anti-BRST transformations, we obtain
\begin{equation}
\begin{align}
 \mathscr{L}_{GF+FP} &= 
  {\alpha+\alpha' \over 2} \mathscr{B} \cdot \mathscr{B} - {\alpha \over 2} ig (\mathscr{C} \times \bar{\mathscr{C}}) \cdot \mathscr{B} 
+ \mathscr{B} \cdot \partial_\mu \mathscr{A}^\mu
\nonumber\\&
 + i \bar{\mathscr{C}} \cdot \partial_\mu \mathscr{D}^\mu[\mathscr{A}]\mathscr{C}
+ {\alpha \over 8}g^2 (\bar{\mathscr{C}} \times \bar{\mathscr{C}}) \cdot (\mathscr{C} \times \mathscr{C}) 
\\
&= {\alpha+\alpha' \over 2}\mathscr{B} \cdot \mathscr{B}
   - {\alpha \over 2} ig (\mathscr{C} \times \bar{\mathscr{C}}) \cdot \mathscr{B} 
+ \mathscr{B} \cdot \partial_\mu \mathscr{A}^\mu 
\nonumber\\&
+ i \bar{\mathscr{C}} \cdot \partial_\mu \mathscr{D}^\mu[\mathscr{A}]\mathscr{C}
+ {\alpha \over 4}g^2 (i \mathscr{C} \times \bar{\mathscr{C}}) \cdot (i \mathscr{C} \times \bar{\mathscr{C}}) ,
\end{align}
\end{equation}
If we impose one more condition, i.e., the FP conjugation invariance,
\begin{equation}
 \mathscr{C}^A \rightarrow \pm \bar{\mathscr{C}}^A, \quad
 \bar{\mathscr{C}}^A \rightarrow \mp \mathscr{C}^A, \quad
 \mathscr{B}^A \rightarrow - \bar{\mathscr{B}}^A, \quad
 \bar{\mathscr{B}}^A \rightarrow - \mathscr{B}^A, \quad
 (\mathscr{A}_\mu^A \rightarrow \mathscr{A}_\mu^A) .
\end{equation}
the second term of (\ref{GFglobal}) is excluded, i.e., $\alpha'=0$.

\par
When $\alpha=0$, this theory reduces to the usual Yang-Mills theory in the Lorentz type gauge fixing with the gauge fixing parameter $\alpha'$:
\begin{equation}
\begin{align}
 \mathscr{L}_{GF+FP} &= 
  {\alpha' \over 2} \mathscr{B} \cdot \mathscr{B} 
+ \mathscr{B} \cdot \partial_\mu \mathscr{A}^\mu + i \bar{\mathscr{C}} \cdot \partial_\mu \mathscr{D}^\mu[\mathscr{A}]\mathscr{C} .
\end{align}
\end{equation}
This is consistent with the FP prescription.
\par
For $\alpha \not=0$, a quartic ghost interaction exists.  Therefore we must go beyond the FP prescription.  Eliminating the Nakanishi-Lautrup field $\mathscr{B}$, we obtain
\begin{equation}
\begin{align}
 \mathscr{L}_{GF+FP} 
&= 
- {1 \over 2\lambda}(\partial_\mu \mathscr{A}^\mu)^2
+ (1-\gamma)i \bar{\mathscr{C}} \cdot \partial_\mu \mathscr{D}^\mu[\mathscr{A}]\mathscr{C}
+ \gamma i \bar{\mathscr{C}} \cdot \mathscr{D}^\mu[\mathscr{A}] \partial_\mu \mathscr{C}
\nonumber\\&
+ {1 \over 2}\lambda \gamma (1-\gamma) g^2 (i \mathscr{C} \times \bar{\mathscr{C}}) \cdot (i \mathscr{C} \times \bar{\mathscr{C}}) ,
\end{align}
\end{equation}
where we have defined
$
  \lambda := \alpha+\alpha'
$
and
$
  \gamma := {\alpha/2 \over \alpha+\alpha'} .
$
When $\gamma=0$ or $\gamma=1$, the quartic ghost interaction disappears and the Lagrangian reduces to the usual FP Lagrangian.
It is shown that $\gamma=0$ and $\gamma=1$ are fixed point of the renormalization group.  Once $\gamma=0$ or $\gamma=1$ is chosen, therefore, $\gamma$ is not renormalized by quantum corrections.  Then we recover the FP Lagrangian.  
The FP conjugation invariance is broken in the usual FP Lagrangian where the ghost and anti-ghost are not treated on equal footing.
In other words, the FP conjutation is recovered for $\alpha'=0$ (or $\gamma=1/2$ and $\lambda=\alpha$) by including the quartic ghost interaction even for $\alpha=0$.

\par
We would like to point out that in the Landau gauge $\alpha=0$ (i.e.,  $\gamma=0$) which was adopted in the works \cite{Boucaud00,GSZ01} claiming the existence of mass dimension 2 gluon condensate, the ghost condensation can not occur.  
There is a fact that the Landau gauge is a fixed point of the theory and hence we do not need to consider the ghost condensation as far as the theory is kept on the Laudau-gauge fixed point.  
This is a reason why we do not need to consider a possibility of the ghost condensation in the Landau gauge supporting the claim of the existence of the gluon condensate (\ref{gluoncond2}) alone.  
In the non-Landau gauges $\alpha\not=0$, the quartic ghost interaction is necessary for renormalizability.  This ghost self-interaction may cause the ghost condensation and hence, even in the Lorentz type gauge fixing, the ghost condensation\begin{equation}
  \langle 0| ig^2 \bar{C}^A C^A |0 \rangle \not= 0 ,
\label{ghostcond2}
\end{equation}
is possible in gauges $\alpha\not=0$ other than the Landau gauge.
In this way, we can understand that two kinds of mass dimension 2 vacuum condensate exist in general in the Yang-Mills theory (and that one of them was claimed to exist earlier in the heuristic works\cite{Schaden99,KS00}),
providing a thorough understanding of this phenomena. 
\par
\par
In the Landau gauge, the OPE of the gluon and ghost propagators have been evaluated, see \cite{Larsson85,LS88,BS89,Spiridonov90,LO92}.
First, the gluon propagator 
$
 D_{\mu\nu}^{AB}(p) := \langle 0|T[\mathscr{A}_\mu^A(p) \mathscr{A}_\nu^B(-p)]|0 \rangle 
$
has the OPE:
\begin{equation}
\begin{align}
  (D_{\mu\nu}^{AB})^{-1}(p) 
  &\sim  C^{[1]}{}_{\mu\nu}^{AB}(p) \langle 0| {\bf 1} |0 \rangle   
  +  C^{[(\mathscr{A}_{\rho}^C)^2]}{}_{\mu\nu}^{AB}(p) \langle 0| (\mathscr{A}_{\rho}^C)^2 |0 \rangle 
  +  C^{[\bar{\mathscr{C}}^C \mathscr{C}^C]}{}_{\mu\nu}^{AB}(p) \langle 0| \bar{\mathscr{C}}^C \mathscr{C}^C |0 \rangle 
\nonumber\\
& +  C^{[(\mathscr{F}_{\rho\sigma}^C)^2]}{}_{\mu\nu}^{AB}(p) \langle 0|(\mathscr{F}_{\rho\sigma}^C)^2 |0 \rangle 
+  C^{[\bar{\mathscr{C}}^C \partial^2 \mathscr{C}^C]}{}_{\mu\nu}^{AB}(p) \langle 0|\bar{\mathscr{C}}^C \partial^2 \mathscr{C}^C |0 \rangle 
+ \cdots ,
\end{align}
\end{equation}
where the global $SU(N)$ invariance is assumed in addition to the Lorentz invariance.
According to the method \cite{Larsson85,TM83},  the Wilson coefficient can be calculated in perturbation theory by equating a $(2+n)$-point one-particle irreducible (1PI) Green's function -- where two of the external legs have hard momentum and the remaining $n$ external legs are assigned zero momentum -- with the coefficient times an $n$-point Green's function with an insertion of the operator under study at zero momentum. 
It should be remarked that the ghost condensate 
$\langle \bar{\mathscr{C}} \mathscr{C} \rangle$
can not appear in this expression, since the ghost--anti-ghost--gluon vertex 
$gf^{ABC}p_\mu$ is proportional to the outgoing ghost momentum $p_\mu$.
Hence we obtain
\begin{equation}
 C^{[\bar{\mathscr{C}}^C \mathscr{C}^C]}{}_{\mu\nu}^{AB}(p) = 0 .
\end{equation}
(This is not the case in the Lorentz gauge with $\alpha\not=0$ and in the MA gauge where the ghost condensation is expected to occur.)
The result of investigations \cite{Larsson85,LS88,Spiridonov90,LO92} shows that the $SU(N_c)$ Yang-Mills theory in the $D$-dimensional spacetime has
\begin{equation}
\begin{align}
 (D_{\mu\nu}^{AB})^{-1}(p) 
 &= \delta^{AB}p^2 [g_{\mu\nu}-(1-\alpha^{-1})p_\mu p_\nu/p^2]
 + \Pi_{\mu\nu}^{AB}(p) ,
\\
  \Pi_{\mu\nu}^{AB}(p) &= {2N_c g^2 \over (N_c^2-1)D} \delta^{AB} 
\left[
  {D-5-\alpha \over 2}  
\langle (\mathscr{A}_\rho^C)^2   \rangle  
+  O\left( 1/p^2 \right)  \right] \left( g_{\mu\nu} - {p_\mu p_\nu \over p^2} \right) .
\end{align}
\end{equation}
In order to satisfy the Slavnov-Taylor identity, the vacuum polarization tensor $\Pi_{\mu\nu}^{AB}$ must be transverse.  
This is indeed the case.
Here, the order $1/p^2$ term includes the ghost condensation 
$\langle \bar{\mathscr{C}}^A \partial^2 \mathscr{C}^A \rangle$  and the gluon condensation 
$\langle (\mathscr{F}_{\mu\nu}^A)^2 \rangle$ of mass dimensions 4.   
\par
Next, the ghost propagator 
$
  (G^{AB})(p) := \langle 0|T[\bar{\mathscr{C}}^A(p) \mathscr{C}^B(-p)]|0 \rangle
$
has the OPE:
\begin{equation}
\begin{align}
  (G^{AB})^{-1}(p)  
  &\cong  C^{[1]}{}^{AB}(p) \langle 0| {\bf 1} |0 \rangle   
  +  C^{[(\mathscr{A}_\rho^C)^2]}{}^{AB}(p) \langle 0| (\mathscr{A}_{\rho}^C)^2 |0 \rangle 
  +  C^{[\bar{\mathscr{C}}^C \mathscr{C}^C]}{}^{AB}(p) \langle 0| \bar{\mathscr{C}}^C \mathscr{C}^C |0 \rangle 
\nonumber\\
& +  C^{[(\mathscr{F}_{\rho\sigma}^C)^2]}{}^{AB}(p) \langle 0|(\mathscr{F}_{\rho\sigma}^C)^2 |0 \rangle 
+  C^{[\bar{\mathscr{C}}^C \partial^2 \mathscr{C}^C]}{}^{AB}(p) \langle 0|\bar{\mathscr{C}}^C \partial^2 \mathscr{C}^C |0 \rangle 
+ \cdots .
\end{align}
\end{equation}
By the same reason as the above, a Wilson coefficient vanishes:
\begin{equation}
  C^{[\bar{\mathscr{C}}^A \mathscr{C}^A]}{}^{AB}(p) = 0 ,
\end{equation}
and hence the ghost condensation $\langle 0| \bar{\mathscr{C}}^C \mathscr{C}^C |0 \rangle$ does not apper in the OPE in this gauge.
The explicit calculation of the Wilson coefficient yields
\begin{equation}
\begin{align}
 (G^{AB})^{-1}(p) &= \delta^{AB}p^2  + \Pi^{AB}(p) ,
\\
  \Pi^{AB}(p) &= {N_c g^2 \over (N_c^2-1)D} \left[
\langle (\mathscr{A}_\mu^A)^2  \rangle  
+   O\left( 1/p^2 \right) \right] .
\end{align}
\end{equation}
Owing to gluon condensation of mass dimension 2, we find that all the gluons acquire the nonvanishing effective mass as
\begin{equation}
  m_A^2 =  {N_c \over (N_c^2-1)D}   
  (\alpha+5-D)  
\langle g^2 \mathscr{A}_\mu^A  \mathscr{A}_\mu^A   \rangle  ,
\end{equation}
whereas all the ghost acquire the effective mass as
\begin{equation}
  m_C^2 = - {N_c \over (N_c^2-1)D} 
\langle g^2 \mathscr{A}_\mu^A  \mathscr{A}_\mu^A   \rangle .
\end{equation}
Remarkably, the nonvanishing mass obtained for $\alpha=0$ in this way does not break the BRST symmetry, since the vacuum polarization tensor is transverse and the ST identity is always satisfied.  
However, the effective mass is gauge parameter dependent.%
\par
Recent numerical simulations suggest the existence of such a $1/p^2$ ultraviolet (UV) correction \cite{Boucaudetal00}.
The numerical simulations suggest the value
$g^2_R \langle \mathscr{A}_\mu^2 \rangle_R = (2.76\text{GeV})^2$ for $SU(3)$.
This value should be compared with the value
$\langle {\alpha \over \pi}(\mathscr{F}_{\mu\nu}^A)^2 \rangle$.
Moreover, the OPE of the three-gluon vertex show that the vacuum condensate of mass dimension 2 yields the UV corrections $\Lambda^2/Q^2$ in the QCD running coupling constant $\alpha_s(Q^2)$:
\begin{equation}
  \alpha_s(Q^2) = \alpha_s(Q^2)_{pert} \left[ 1+{g_R^2\langle \mathscr{A}_\mu^2 \rangle_R \over 4(N_c^2-1)}{9 \over Q^2} +O(\alpha) \right] .
\end{equation}
Such a correction for the coupling constant \cite{AZ98,BRMO98} leads to a piece of the linear potential 
$\sigma_s r$
at short distances $r \rightarrow 0$
in addition to the usual Coulomb-like potential $-C_F {\alpha_s(r) \over r}$ with
$C_F={N_c^2-1 \over 2N_c}$.  Here the string tension of the short-distance linear potential is proportional to the the gluon condensate of mass dimension 2:
\begin{equation}
  \sigma_s \cong g_R^2\langle \mathscr{A}_\mu^2 \rangle_R .
\end{equation}
The right hand side of this equation is BRST and anti-BRST invariant (A special case $\alpha=0$ of the operator $\mathcal{O}$). So is the left hand side, i.e., the string tension of the short string.
This argument suggests that the quark confinement is already intertwined in the UV region of QCD. 
Thus it is expected that the linear potential or the area law of the Wilson loop at short distances $r \ll 1$ could well be extrapolated to large distances $r \gg 1$ by including higher-order corrections up to a desired order.
\footnote{
In the course of preparing this paper, we noticed that the value of the gluon condensate of mass dimension 2 in the Landau gauge of the Lorentz type covariant gauge has been evaluated by Verschelde et al.\cite{VKAV01} by making use of the multiplicative renormalizable effective potential for this condensate.
}
In the case of $\alpha\not=0$, it is easy to understand (by repeating the OPE calculation to the lowest non-trivial order $g^2$) that the quartic ghost interaction induces the ghost condensation of mass  dimension 2 in the OPE and hence shifts the effective mass by this contribution.  However, it is not yet checked whether the combination of two condensations of mass dimension 2 obtained from the OPE agrees with the BRST and anti-BRST invariant combination (\ref{combi}).
In this sense, the Landau gauge is the most economical gauge in the Lorentz type gauge fixing, since we do not need to worry about this issue.

\subsection{The modified MA gauge}
\par
Now we turn our attention to the modified MA gauge \cite{KondoII},
\begin{equation}
 \mathscr{L}_{GF+FP} = i \bm{\delta}_B \bar{\bm{\delta}}_B \left[ {1 \over 2}A_\mu^a(x) A^{\mu}{}^a(x)
 - {\alpha \over 2}i C^a(x) \bar{C}^a(x) \right] .
\label{MAGF}
\end{equation}
This is rewritten as
\begin{equation}
\begin{align}
  \mathscr{L}_{GF+FP}' &=     
{\alpha \over 2} B^a B^a
+ B^a D_\mu[a]^{ab}A^\mu{}^b
- \alpha  g f^{abi} i B^a \bar C^b C^i 
+ {\alpha \over 2} g f^{abc} i B^b C^a \bar C^c  
\nonumber\\
&+ i \bar C^a D_\mu[a]^{ac} D^\mu[a]^{cb} C^b
- i g^2 f^{adi} f^{cbi} \bar C^a C^b A^\mu{}^c A_\mu^d 
\nonumber\\
&+ i \bar C^a D_\mu[a]^{ac}(g f^{cdb}  A^\mu{}^d C^b)
+ i \bar C^a g  f^{abi} (D^\mu[a]^{bc}A_\mu^c) C^i 
\nonumber\\
&+{\alpha \over 8} g^2 f^{abe}f^{cde} \bar C^a \bar C^b C^c C^d
+ {\alpha \over 4} g^2 f^{abc} f^{aid} \bar C^b \bar C^c C^i C^d
\nonumber\\
&
+ {\alpha \over 4} g^2 f^{abi} f^{cdi} \bar C^a \bar C^b C^c C^d \} .
\label{GF3}
\end{align}
\end{equation}
In particular, the $SU(2)$ case is greatly simplified as ($a,b,c,d=1,2$)
\begin{equation}
\begin{align}
  \mathscr{L}_{GF+FP}' &=   
 {\alpha \over 2} B^a B^a
+ B^a D_\mu[a]^{ab}A^\mu{}^b
- \alpha  g \epsilon^{ab} i B^a \bar C^b C^3
\nonumber\\
&+ i \bar C^a D_\mu[a]^{ac} D^\mu[a]^{cb} C^b
- i g^2 \epsilon^{ad} \epsilon^{cb} \bar C^a C^b A^\mu{}^c A_\mu^d 
\nonumber\\
&
+ i \bar C^a g  \epsilon^{ab} (D_\mu[a]^{bc}A_\mu^c) C^3 
\nonumber\\
& 
+ {\alpha \over 4} g^2 \epsilon^{ab} \epsilon^{cd} \bar C^a \bar C^b C^c C^d .
\label{MAGF}
\end{align}
\end{equation}
A characteristic feature of the MA gauge is that it breaks the global $SU(N)$ symmetry explicitly as well as the local $SU(N)$ symmetry.  However, the global gauge symmetry is broken by the BRST-exact term, since the breaking is caused by the gauge fixing plus Faddeev-Popov  term of the form, (\ref{MAGF}).  In this sense, the MA gauge resembles the renormalizable $R_\xi$ gauge in the gauge-Higgs theory where  the would-be Nambu-Goldstone mode in the GF+FP term breaks explicitly the original global gauge symmetry and hence the massless Nambu-Golstone particle does not appear.
For more details of the modified MA gauge, see \cite{KondoII,Kondo00,KondoIII,KondoIV,KondoV,KondoVI,KS00,KS01,SIK01,MLP85,HN93}.
\par
In the $SU(2)$ Yang-Mills theory in the $D$-dimensional spacetime, the  OPE of the propagator in the modified MA gauge is obtained as follows 
(The full details including higher order terms will be given in a forthcoming paper \cite{KIMS01}).
 According to the standard method \cite{Larsson85,TM83} and the Feynman rules\cite{KS01,SIK01}, it is shown that the off-diagonal gluon propagator 
$
 D_{\mu\nu}^{ab}(p) := \langle 0|T[A_\mu^a(p) A_\nu^b(-p)]|0 \rangle 
$
has the OPE:
\begin{equation}
\begin{align}
 &  (D_{\mu\nu}^{ab})^{-1}(p)  
\nonumber\\
&\sim  C^{[1]}{}_{\mu\nu}^{ab}(p) \langle 0| {\bf 1} |0 \rangle  
  +  C^{[(A_{\rho}^c)^2]}{}_{\mu\nu}^{ab}(p) \langle 0| (A_{\rho}^c)^2 |0 \rangle 
 +  C^{[\bar{C}^c C^c]}{}_{\mu\nu}^{ab}(p) \langle 0|\bar{C}^c C^c |0 \rangle 
+ \cdots 
 \nonumber\\
 &= \delta^{ab}p^2 [g_{\mu\nu}-(1-\alpha^{-1})p_\mu p_\nu/p^2]
 +   i \Pi_{\mu\nu}^{ab}(p) ,
\\  
& i \Pi_{\mu\nu}^{ab}(p)  
\nonumber\\
&=   \delta^{ab} {g^2 \over 2D}
\left\{ 2 \left(  D-5-\beta \right)  g_{\mu\nu}
+ \left[ 4-\beta \left( {2 \over \alpha}+2 \right) - \left( 1+{1 \over \alpha} \right)^2(D-1+\beta) \right] {p_\mu p_\nu \over p^2}  \right\} 
\langle  (A_\rho^c)^2 \rangle 
\nonumber\\
&+ \delta^{ab}  g^2 g_{\mu\nu}    
\langle i \bar{C}^c C^c \rangle  
+ O\left( 1/p^2 \right) ,
\end{align}
\end{equation}
where $\beta$ is the gauge fixing parameter of the Lorentz type gauge fixing for the diagonal gluon.
It is shown that 
$C^{[\epsilon^{cd}\bar{C}^c C^d]}{}_{\mu\nu}^{ab}(p)=0$ up to $O(g^2)$.  Hence the mixed off-diagonal condensation $\langle i\epsilon^{cd} \bar{C}^c C^d \rangle$ does not appear in the OPE and hence in the effective mass of the off-diagonal gluons at least to this order.%
\footnote{
The mixed ghost condensation has a nonvanishing value 
$\langle ig^2 \epsilon^{cd} \bar{C}^c C^d \rangle\not=0$
and breaks the global $SL(2,R)$ symmetry in the $SU(2)$ Yang-Mills theory, see \cite{Schaden99,KS00}.
}
This result agrees with that of the effective potential approach which will be reported later \cite{KIMS01}.
In the MA gauge, the vacuum polarization of the off-diagonal gluon is not transverse, whereas that of the diagonal-gluon is transverse owing to the residual U(1) symmetry.  In other words, the off-diagonal gluon field $A_\mu^a$ and the gauge fixing parameter $\alpha$ receive the different renormalization, as explicitly evaluated in \cite{Schaden99} for $SU(2)$ and \cite{KS01} for $SU(N)$ at one-loop level.
\par
The off-diagonal ghost propagator 
$
  (G^{ab})(p) := \langle 0|T[\bar{C}^a(p) C^b(-p)]|0 \rangle
$
has the OPE:
\begin{equation}
\begin{align}
  (G^{ab})^{-1}(p)  &\sim  C^{[1]}{}^{ab}(p) \langle 0| {\bf 1} |0 \rangle  
  +  C^{[(A_{\rho}^c)^2]}{}^{ab}(p) \langle 0| (A_{\rho}^c)^2 |0 \rangle 
 +  C^{[\bar{C}^c C^c]}{}^{ab}(p) \langle 0|\bar{C}^c C^c |0 \rangle 
+ \cdots 
 \nonumber\\
 &= \delta^{ab}p^2   + \Pi^{ab}(p) ,
\\
 \Pi^{ab}(p) &=  - \langle g^2 (A_\rho^c)^2 \rangle \delta^{ab}
+ \left( -{1 \over 2}\alpha+\beta \right) \langle ig^2 \bar{C}^c C^c \rangle \delta^{ab} 
+ O\left( 1/p^2 \right) .
\end{align}
\end{equation}

\par
Thus we can identify the effective mass of the off-diagonal gluon 
with a linear combination of two vacuum condensates of mass dimension 2 as
\begin{equation}
  m_A^2 =  {1+\beta \over 4} \langle g^2 A_\mu^a A_\mu^a \rangle 
+ \langle ig^2 \bar{C}^a C^a \rangle  .
\end{equation}
Similarly, the off-diagonal ghost has the effective mass given by
\begin{equation}
  m_C^2 =  \langle g^2 A_\mu^a A_\mu^a \rangle 
+ \left( {1 \over 2}\alpha+\beta \right) \langle ig^2 \bar{C}^a C^a \rangle  .
\end{equation}
These values should be compared with those in the previous work \cite{KS00}.
\par
Although we can give the arguments which are similar to those in the Lorentz gauge given above, we do not repeat them here.  Nevertheless, we shall give a comment on the relationship of this result with quark confinement.
In the previous paper \cite{Kondo00}, we have claimed that the string tension is proportional to the off-diagonal gluon mass (up to a logarithmic correction):
\begin{equation}
  \sigma_{st} \cong m_A^2 .
\label{st}
\end{equation}
If the effective mass of the off-diagonal gluon was extrapolated to the infrared (IR) region so the the pole mass was obtained  in the gauge invariant combination, i.e., $m_A^2 \sim g^2 \langle \mathcal{O} \rangle$ from the proposed operator $\mathcal{O}$ of mass dimension 2, the above relation (\ref{st}) could lead to the BRST invariant string tension in the gauge fixed formulation of the Yang-Mills theory.
The RG approach to this issue will be given elsewhere.

\section{Conclusion and discussion}
\par
In this paper we have investigated a possibility of gluon and ghost condensation of mass dimension 2 in the Yang-Mills theory with the manifestly covariant gauge fixings, i.e., the Lorentz gauge and the modified MA gauge.
We have discussed that the condensation of mass dimension 2 provides a mechanism of effective mass generation of the gluons and ghosts by demonstrating their existence in the OPE of the gluon and ghost propagators.
Especially, we have proposed a BRST-invariant combination of two vacuum condensates of mass dimension 2, which reduces in the Landau gauge $\alpha=0$ of the Lorentz gauge to the gluon condensates of mass  dimension 2 proposed recently by several authors\cite{Boucaud00,GSZ01} and include the ghost condensation in the modified MA gauge claimed in \cite{Schaden99,KS00}.
\par
However, the gluon and ghost propagators are not BRST invariant.  Therefore, there is no guarantee as to whether a linear combination of the vacuum condensates derived by the OPE can be BRST invariant or not.
The RG approach can connect the UV region to the infrared (IR) region by successive integration of high-energy modes in a systematic way.
Anyway, this point clearly warrants further studies. 
\par
The existence of mass dimension 2 vacuum condensate can shed more light on clarifying the mechanism of color confinement from a nonperturbative point of view.
The interplay between the chiral symmetry breaking and the quark confinement can also be discussed by including the quarks into the above formalism.  
The OPE result obtained in this paper can provide  a good starting point for the Schwinger-Dyson equation approach to clarify a non-perturbative features of QCD.  


\appendix
\section{BRST invariance of the proposed operator $\mathcal{O}$}
\setcounter{equation}{0}

For the Yang-Mills gauge theory, the BRST transformation is given by
\begin{subequations}
\begin{equation}
\begin{align}
 \bm{\delta}_B \mathscr{A}_\mu(x) &=
\mathscr{D}_\mu[\mathscr{A}]\mathscr{C}(x) :=
 \partial_\mu \mathscr{C}(x) + g (\mathscr{A}_\mu(x) \times \mathscr{C}(x)) , \\
 \bm{\delta}_B \mathscr{C}(x) &= -{1 \over 2}g(\mathscr{C}(x) \times \mathscr{C}(x)) ,
\\
 \bm{\delta}_B \bar{\mathscr{C}}(x) &= i \mathscr{B}(x) ,
\\
 \bm{\delta}_B \mathscr{B}(x) &= 0 ,
\end{align}
\label{BRST1}
\end{equation}
\end{subequations}
where%
$\mathscr{A}_\mu, \mathscr{B}, \mathscr{C}$ and $\bar{\mathscr{C}}$ are the gauge field, the Nakanishi-Lautrap (NL) auxiliary field, the Faddeev--Popov (FP) ghost and anti-ghost fields respectively.
Another BRST transformation, say anti-BRST transformation~\cite{antiBRST}), is  given by
\begin{subequations}
\begin{equation}
\begin{align}
 \bar{\bm{\delta}}_B \mathscr{A}_\mu(x) &= 
\mathscr{D}_\mu[\mathscr{A}] \bar{\mathscr{C}}(x) :=
\partial_\mu \bar{\mathscr{C}}(x) + g (\mathscr{A}_\mu(x) \times \bar{\mathscr{C}}(x)) , \\
 \bar{\bm{\delta}}_B \bar{\mathscr{C}}(x) &= -{1 \over 2}g(\bar{\mathscr{C}}(x) \times \bar{\mathscr{C}}(x)) ,
 \\
 \bar{\bm{\delta}}_B \mathscr{C}(x) &= i \bar{\mathscr{B}}(x) ,
\\
 \bar{\bm{\delta}}_B \bar{\mathscr{B}}(x) &= 0 ,
\end{align}
\label{BRST2}
\end{equation}
\end{subequations}
where $\bar{\mathscr{B}}$ is defined by
\begin{equation}
  \bar{\mathscr{B}}(x) = -\mathscr{B}(x) + ig (\mathscr{C}(x) \times \bar{\mathscr{C}}(x)) .
\end{equation}
Here we have used the following notation:
\begin{equation}
 F \cdot G := F^A G^A, \quad 
 F^2 := F \cdot F , \quad 
 (F \times G)^A := f^{ABC}F^B G^C ,
\end{equation}
where $f^{ABC}$ are the structure constants of the Lie algebra $\mathscr{G}$ of the gauge group $G$.

\subsection{Lorentz gauge}
We consider only the $\alpha'=0$ case.
By eliminating the Nakanishi-Lautrap field $\mathscr{B}$, the on-shell BRST and anti-BRST transformations are obtained as
\begin{equation}
\begin{align}
 \bm{\delta}_B \bar{\mathscr{C}}(x) 
 &=  -{i \over \alpha} \partial^\mu \mathscr{A}_\mu(x) - {1 \over 2}g \mathscr{C}(x) \times \bar{\mathscr{C}}(x)   ,
\\
 \bar{\bm{\delta}}_B \mathscr{C}(x) 
&=  +{i \over \alpha} \partial^\mu \mathscr{A}_\mu(x) - {1 \over 2}g \mathscr{C}(x) \times \bar{\mathscr{C}}(x)    .
\end{align}
\label{BRSTonshell}
\end{equation}
The BRST transformation of the operator $\mathcal{O}$ is calculated as
\begin{equation}
\begin{align}
  \bm{\delta}_B \mathcal{O} &=  (\Omega^{(D)})^{-1} \int d^D x \  \bm{\delta}_B \left[ 
 {1 \over 2} \mathscr{A}_\mu^A(x) \mathscr{A}^\mu{}^A(x) - \alpha i \mathscr{C}^A(x)   \bar{\mathscr{C}}^A(x) \right] 
\nonumber\\
 &=  (\Omega^{(D)})^{-1} \int d^D x \   \left[ 
  \mathscr{A}_\mu^A(x) \bm{\delta}_B \mathscr{A}^\mu{}^A(x) - \alpha i \bm{\delta}_B \mathscr{C}^A(x)   \bar{\mathscr{C}}^A(x)
+ \alpha i  \mathscr{C}^A(x)   \bm{\delta}_B \bar{\mathscr{C}}^A(x) \right] 
\nonumber\\
&= (\Omega^{(D)})^{-1} \int d^D x \partial^\mu (\mathscr{A}_\mu^A(x) \mathscr{C}^A(x)) .
\end{align}
\end{equation}
The anti-BRST invariance is also shown in the similar way.
Moreover, the vacuum expectation value of $\bm{\delta}_B \mathcal{O}$ is zero owing to the translational invariance of the vacuum,
\begin{equation}
\begin{align}
  \langle 0| \bm{\delta}_B \mathcal{O} |0 \rangle  
= \langle 0| [iQ_B, \mathcal{O}] |0 \rangle 
= (\Omega^{(D)})^{-1} \int d^D x \partial^\mu 
\langle 0| \mathscr{A}_\mu(x) \cdot \mathscr{C}(x) |0 \rangle  = 0 ,
\end{align}
\end{equation}
in agreement with $Q_B|0 \rangle=0$.

\subsection{Modified MA gauge}
In the modified MA gauge, the on-shell BRST and anti-BRST transformation is given by
\begin{equation}
\begin{align}
 \bm{\delta}_B \bar{C}^a(x) 
 &= 
- {i \over \alpha}D_\mu^{ab}[a(x)]A^\mu{}^b(x)   
+ ig f^{abi} i \bar{C}^b(x) C^i(x) 
+ {i \over 2}g f^{abc}i C^b(x) \bar{C}^c(x) ,
\\
 \bar{\bm{\delta}}_B C^a(x) 
&= + {i \over \alpha}D_\mu^{ab}[a(x)]A^\mu{}^b(x)   
+ ig f^{abi} i C^b(x) \bar{C}^i(x) 
+ {i \over 2}g f^{abc}i C^b(x) \bar{C}^c(x) ,
\end{align}
\label{BRSTonshell2}
\end{equation}
where
$
  D_\mu^{ab}[a] := \partial_\mu \delta^{ab} -g f^{abi}a_\mu^i .
$
If we calculate the BRST transformation of the operator $\mathcal{O}$, all the trilinear terms in the field coming from the BRST transformation cancel to obtain a simple result,
\begin{equation}
\begin{align}
  \bm{\delta}_B \mathcal{O} &=  (\Omega^{(D)})^{-1} \int d^D x \  \bm{\delta}_B \left[ 
 {1 \over 2} A_\mu^a(x) A^\mu{}^a(x) - \alpha i C^a(x)   \bar{C}^a(x) \right] 
\nonumber\\
&= (\Omega^{(D)})^{-1} \int d^D x \partial^\mu (A_\mu^a(x) C^a(x)) .
\end{align}
\end{equation}



\baselineskip 14pt

\begin{thebibliography}{99}
\bibitem{SVZ79}
  M.A. Shifman, A.I. Vainshtein and V.I. Zakharov,
QCD and resonance physics. theoretical foundations,
Nucl. Phys. B 147, 385-447 (1979).
\\
  M.A. Shifman, A.I. Vainshtein and V.I. Zakharov,
QCD and resonance physics. applications,
Nucl. Phys. B 147, 448-518 (1979).

\bibitem{Schaden99}
  M. Schaden,
  Mass generation in continuum SU(2) gauge theory in covariant Abelian gauges,
  hep-th/9909011, 3rd revised version.

\bibitem{KS00}
  K.-I. Kondo and T. Shinohara,
  Abelian dominance in low-energy Gluodynamics due to dynamical mass generation,
  hep-th/0004158,
  Phys. Lett. B 491, 263-274 (2000).

\bibitem{Kondo00}
  K.-I. Kondo,
  Dual superconductivity, monopole condensation and confining string in low-energy Yang-Mills theory,
CHIBA-EP-123,
hep-th/0009152 (revised version in preparation).

\bibitem{KondoI}
  K.-I. Kondo,
  Abelian-projected effective gauge theory of QCD with asymptotic
freedom and quark confinement,
  hep-th/9709109,
  Phys. Rev. D 57, 7467-7487 (1998).
  \\
K.-I. Kondo, 
  hep-th/9803063,
  Prog. Theor. Phys. Supplement, No. 131, 243-255.


\bibitem{KondoII}
  K.-I. Kondo, 
  Yang-Mills theory as a deformation of topological field
theory, dimensional reduction and quark confinement,
  hep-th/9801024,
  Phys. Rev. D 58, 105019 (1998).

\bibitem{dualsuper}
  Y. Nambu,
  Strings, monopoles, and gauge fields,
  Phys. Rev. D 10, 4262-4268 (1974).
\\
G. 't Hooft,
  in: High Energy Physics, edited by A. Zichichi 
(Editorice Compositori, Bologna, 1975).
\\
S. Mandelstam,
  Vortices and quark confinement in non-abelian gauge theories, 
 Phys. Report  23, 245-249 (1976).
\\
A.M. Polyakov,
  Compact gauge fields and the infrared catastrophe,
  Phys. Lett. B 59, 82-84 (1975).
  Quark confinement and topology of gauge theories,
  Nucl. Phys. B 120, 429-458 (1977).

\bibitem{tHooft81}
  G. 't Hooft,
  Topology of the gauge condition and new confinement
phases in non-Abelian gauge theories,
  Nucl.Phys. B 190 [FS3], 455-478 (1981).

\bibitem{EI82}
  Z.F. Ezawa and A. Iwazaki,
  Abelian dominance and quark confinement in Yang-Mills
theories,
  Phys. Rev. D 25, 2681-2689 (1982).

\bibitem{SY90}
  T. Suzuki and I. Yotsuyanagi,
  Possible evidence of abelian dominance in quark
confinement,
  Phys. Rev. D 42, 4257-4260 (1990).


\bibitem{monopole}
S. Hioki, S. Kitahara, S. Kiura, Y. Matsubara, O. Miyamura, S. Ohno, T. Suzuki,
Phys. Lett. B 272, 326-332 (1991), Erratum-ibid. B281, 416 (1992). 
\\
J.D. Stack, S.D. Neiman and R. Wensley,
hep-lat/9404014,
Phys. Rev. D 50, 3399-3405 (1994).
\\
H. Shiba and T. Suzuki,
hep-lat/940401,
Phys. Lett. B 333, 461-466 (1994).

\bibitem{Boucaud00}
  Ph. Boucaud, A. Le Yaouanc, J.P. Leroy, J. Micheli, O. Pene and J. Rodriguez-Quintero,
Consistent OPE description of gluon two- and three-point Green functions?,
hep-ph/0008043,
Phys. Lett. B 493, 315-324 (2000).


\bibitem{GSZ01}
  F.V. Gubarev, L. Stodolsky and V.I. Zakharov,
On the significance of the vector potential squared,
hep-th/0010057,
Phys. Rev. Lett. 86, 2220-2222 (2001).


\bibitem{GZ01}
  F.V. Gubarev and V.I. Zakharov,
Emerging phenomenology of $\langle A^2_{min} \rangle$,
hep-ph/0010096,
Phys. Lett. B 501, 28-36 (2001).

\bibitem{Wilson69}
  K.G. Wilson,
Non-Lagrangian models of current algebra,
  Phys. Rev. 179, 1499-1512 (1969).

\bibitem{WK74}
  K.G. Wilson and J. Kogut,
  The renormalization group and the $\epsilon$ expansion,
Phys. Report 12, 75-200 (1974).

\bibitem{KIMS01}
  K.-I. Kondo, T. Imai, T. Murakami and T. Shinohara,
in preparation.

\bibitem{BRST}
  C. Becchi, A. Rouet and R. Stora,
Renormalization of the Abelian Higgs model,
  Commun. Math. Phys. 42, 127-162 (1975);
Renormalization of gauge theories,
  Ann. Phys. 98, 287-321 (1976).
\\
I.V. Tyutin,
  Lebedev preprint, FIAN No.39 (in Russian) (1975).

\bibitem{antiBRST}
  G. Curci and R. Ferrari,
  Slavnov transformations and supersymmetry,
  Phys. Lett. B 63, 91-94 (1976).
  \\
I. Ojima,
  Another BRS transformation,
  Prog. Theor. Phys. 64, 625-638 (1980).



\bibitem{BT82}
  L. Baulieu and J. Thierry-Mieg,
The principle of BRS symmetry: An alternative approach to Yang-Mills theories,
Nucl. Phys. B 197, 477-508 (1982).

\bibitem{Baulieu85}
L. Baulieu,
Perturbative gauge theories,
Phys. Reports 129, 1-74 (1985).

\bibitem{DJ82}
  R. Delbourgo and P.D. Jarvis,
Extended BRS invariance and OSP(4/2) supersymmetry,
J. Phys. A: Math. Gen. 15, 611-625 (1982).





\bibitem{Larsson85}
  T.I. Larsson,
Nonperturbative propagators in quantum chromodynamics,
Phys. Rev. D32, 956-961 (1985).

\bibitem{LS88}
  M.J. Lavelle and M. Schaden,
Propagators and condensation in QCD,
Phys. Lett. B 208, 297-302 (1988).

\bibitem{BS89}
  E. Bagan and T.G. Steele,
Are ghost condensates necessary for the Slavnov-Taylor identities?,
Phys. Lett. B 226, 142 (1989).

\bibitem{Spiridonov90}
  V.P. Spiridonov,
Operator product expansions, Slavnov-Taylor identities and $d=4$ condensates,
Mod. Phys. Lett. A 5, 653-660 (1990).

\bibitem{LO92}
  M. Lavelle and M. Oleszczuk,
The operator product expansion of the QCD propagators,
Mod. Phys. Lett. A 7, 3617-3630 (1992).

\bibitem{TM83}
  C. Taylor and B. McClain,
Operator-product expansion and the asymptotic behavior of spontaneously broken scalar field theories,
Phys. Rev. D28, 1364-1371 (1983).




\bibitem{Boucaudetal00}
  P. Boucaud, G. Burgio, F. Di Renzo, J.P. Leroy, J. Micheli, C. Parrinello, O. Pene, C. Pittori, J. Rodriguez-Quintero, C. Roiesnel and K. Sharkey, 
Lattice calculation of $1/p^2$ corrections to $\alpha_s$ and of $\Lambda_{QCD}$ in the MOM scheme,
hep-ph/0003020,
JHEP 04, 006 (2000).

\bibitem{AZ98}
  R. Akhoury and V.I. Zakharov,
On nonperturbative corrections to the potential for heavy quarks,
hep-ph/9710487,
Phys. Lett. B 438, 165-172 (1998).

\bibitem{BRMO98}
  G. Burgio, F. Di Renzo, G. Marchesini and E. Onofri,
$\Lambda^2$-contribution to the condensate in lattice gauge theory,
Phys. Lett. B 422, 219-226 (1998).


\bibitem{VKAV01}
  H. Verschelde, K. Knecht, K. Van Acoleyen and M. Vanderkelen,
The non-perturbative groundstate of QCD and the local composite operator $A_\mu^2$,
hep-th/0105018.








\bibitem{KondoIII}
  K.-I. Kondo,
  Existence of confinement phase in quantum electrodynamics,
  hep-th/9803133,
  Phys. Rev. D 58, 085013 (1998).

\bibitem{KondoIV}
  K.-I. Kondo,
  Abelian magnetic monopole dominance in quark confinement,
  hep-th/9805153,
  Phys. Rev. D 58, 105016 (1998).

\bibitem{KondoV}
  K.-I. Kondo,
  Quark confinement and deconfinement in QCD from the viewpoint of
Abelian-projected effective gauge theory,
  hep-th/9810167,
  Phys. Lett. B 455, 251-258 (1999).

\bibitem{KondoVI}
  K.-I. Kondo,
  A formulation of the Yang-Mills theory as deformation of a topological field 
  hep-th/9904045,
  Intern. J. Mod. Phys. A 16, 1303-1346 (2001).


\bibitem{KS01}
K.-I. Kondo and T. Shinohara,
  Renormalizable Abelian-projected effective gauge theory derived from quantum chromodynamics,
  hep-th/0005125,
Prog. Theor. Phys. 105, 649-665 (2001).
\\
  T. Shinohara,
  Renormalizable Abelian-projected effective gauge theory derived from quantum chromodynamics. II,
  Chiba Univ. Preprint, CHIBA-EP-127,
  hep-th/0105262.
%
\bibitem{SIK01}
  T. Shinohara, T. Imai and K.-I. Kondo,
The most general and renormalizable maximal Abelian gauge,
CHIBA-EP-128, hep-th/0105268.

\bibitem{MLP85}
  H. Min, T. Lee and P.Y. Pac,
  Renormalization of Yang-Mills theory in the abelian gauge,
  Phys. Rev. D 32, 440-449 (1985).

\bibitem{HN93}
  H. Hata and I. Niigata,
  Color confinement, abelian gauge and renormalization group flow,
  Nucl. Phys. B 389, 133-152 (1993).





\end{thebibliography}
\end{document}